\documentstyle[myproceeding]{article}

\input{epsf.sty}
\def\mydate{10 March 2003}
\def\ignore#1{{}}

\newcommand{\beeq}{\begin{equation}}
\newcommand{\eneq}{\end{equation}}
\newcommand{\beqn}{\begin{eqnarray}}
\newcommand{\eeqn}{\end{eqnarray}}

\def\mybig{\displaystyle \strut }

\def\dd{\partial}
\def\la{\raise.16ex\hbox{$\langle$}\lower.16ex\hbox{}  }
\def\ra{\, \raise.16ex\hbox{$\rangle$}\lower.16ex\hbox{} }
\def\go{\rightarrow}

\def\onehalf{ \hbox{${1\over 2}$} }

\def\Tr{{\rm Tr \,}}

\def\eff{{\rm eff}}
\def\H{{\mathcal H}}

\def\sym{{\rm sym}}

\def\diag{{\rm diag ~}}
\def\BC{{\rm BC}}

\def\coverM{{{\mathcal M}_{\rm cover}}}

\def\psibar{ \psi \kern-.65em\raise.5em\hbox{$-$} }
\def\psibarl{ \psi \kern-.65em\raise.6em\hbox{$-$} \lower.6em\hbox{} }

\def\mycite#1{\cite{#1}${}^{, \,\,}$}

\begin{document}

{\small \noindent \mydate  \hfill OU-HET 432/2003}
\vskip 1.5cm 

\title{GUT ON ORBIFOLDS\footnote{~To appear in the Proceedings of 
the 2002 International Workshop  on {\it ``Strong Coupling Gauge Theories 
and Effective Field Theories''}, Nagoya, Japan, 10-13 December 2002.}\\
-- Dynamical Rearrangement of Gauge Symmetry --}

\author{Yutaka HOSOTANI}

\address{Department of Physics, Osaka University,
Toyonaka, Osaka 560-0043, Japan\\
E-mail: hosotani@phys.sci.osaka-u.ac.jp}

\maketitle
\abstracts{
Grand unified theory defined on higher-dimensional
orbifolds  provides a new way to solve the  hierarchy problem.
In gauge theory on an orbifold  many different sets of
boundary conditions imposed at orbifold fixed points (branes) 
are related by  large gauge transformations,  forming 
an equivalent class of boundary conditions.  Thanks to the Hosotani
mechanism    the physics remains the same in all theories
in a given  equivalent class, though the symmetry of boundary conditions
differs from each other.  Quantum dynamics of Wilson line phases rearranges
the gauge symmetry.  In the nonsupersymmetric
$SU(5)$ model the presence of bulk fermions leads to the spontaneous
breaking of color  $SU(3)$.  In the supersymmetric  model with 
Scherk-Schwarz SUSY breaking,  color $SU(3)$ is spontaneously broken
even in the absence of bulk fermions if there are Higgs multiplets.
}

\section{Introduction}

There  are good reasons for investigating higher dimensional gauge 
theories.  If the superstring theory is to describe the nature, 
we live in ten-dimensional spacetime.    There must be hidden dimensions
beyond the four-dimensional spacetime we see at the current energy scale.
Dynamics in the string theory may result  spacetime with
the structure of an orbifold such that   the
four-dimensional  spacetime we see be located at its fixed points
(brane).  The presence of D-branes in the string theory makes it quite
probable that an effective gauge theory emerges in more than four
dimensions.

Another important reason lies in the fact  that many puzzles or problems
difficult to  be solved in four-dimensional theory can be naturally
resolved thanks to the existence of extra dimensions and the special nature
of orbifolds.\mycite{Hall1}\cite{Quiros1}  Among  the notable problems
are the hierarchy problem in grand unified theories and the chiral fermion
problem.  

In formulating gauge theory on an orbifold, however, there appears,
at a first look,  arbitrary choice of boundary conditions imposed
on   fields at orbifold fixed points (branes).  
This  arbitrariness poses a 
serious obstacle to constructing unified theories in a convincing manner.

In this paper we shall show that the arbitrariness problem in
the orbifold boundary conditions is partially solved at the quantum level
by the  Hosotani 
mechanism.\mycite{Hosotani1}\mycite{Hosotani2}\cite{HHHK} 
Dynamics of Wilson line phases play a vital role in rearranging the gauge
symmetry.  Physical symmetry, in general, differs from the symmetry of the
orbifold boundary conditions.  Physical symmetry does not depend on the
orbifold boundary conditions so long as the boundary conditions belong to
the same equivalence class, whose qualification  shall be detailed below.

\section{Orbifold  boundary conditions in gauge theory}

We shall consider gauge theories on 
${\mathcal M} = M^4 \times (S^1/Z_2)$ where $M^4$ is the four-dimensional
Minkowski spacetime.    Let $x^\mu$ and $y$ be coordinates of
$M^4$ and $S^1$, respectively.   $S^1$ has a radius $R$ so that
a point $(x^\mu, y+2\pi R)$ is identified with a point $(x^\mu, y)$.
The orbifold $M^4 \times (S^1/Z_2)$ is obtained by further identifying
$(x^\mu, -y)$ and $(x^\mu, y)$.  Gauge fields and Higgs fields are defined
in the five dimensional spacetime ${\mathcal M}$.  We adopt  the brane picture
in which quarks and leptons are confined in one of the boundary branes,
say,  at $y=0$.  There arises no problem in having chiral fermions
on the brane.  It is possible to have fermions in the bulk five dimensional
 ${\mathcal M}$, whose effect in the context of the Hosotani mechanism is also
evaluated. 

Fields are defined on the covering space $\coverM$ of $M^4 \times S^1$. 
Physical quantities must be single-valued after a loop translation along
$S^1$ and after $Z_2$ parity reflection around $y=0$ or $y=\pi R$.
This, however, does not imply the single-valuedness of the fields.  
In gauge theory the fields need to return to their original values
up to a gauge transformation and a sign.   Take the
lower four-dimensional components of the gauge potentials, $A_\mu(x,y)$.
They satisfy
\beqn
&&\hskip -1cm 
A_\mu (x, - y ) = P_0 A_\mu(x,y) P_0^\dagger \cr
\noalign{\kern 2pt}
&&\hskip -1cm 
A_\mu (x, \pi R - y ) =  P_1 A_\mu (x,\pi R + y) P_1^\dagger \cr
\noalign{\kern 2pt}
&&\hskip -1cm 
A_\mu (x,  y + 2\pi R ) =  U A_\mu (x,   y) U^\dagger   
\label{BC1} 
\eeqn
where $P_0^2 = P_1^2 = 1$, $P_0^\dagger=P_0$,  $P_1^\dagger=P_1$, 
and $U U^\dagger = 1$.   There holds a relation $U = P_1 P_0$.  

It follows from (\ref{BC1}) that 
$F_{\mu\nu}(x,-y) = P_0 F_{\mu\nu}(x,y) P_0^\dagger$.  The gauge 
covariance can be maintained for the $\mu$-$y$ component only if 
$F_{\mu y}(x,-y) = - P_0 F_{\mu y}(x,y) P_0^\dagger$ etc., which implies
\beqn
&&\hskip -1cm 
A_y (x, - y ) = - P_0 A_y(x,y) P_0^\dagger \cr
\noalign{\kern 2pt}
&&\hskip -1cm 
A_y (x, \pi R - y ) = -  P_1 A_y (x,\pi R + y) P_1^\dagger \cr
\noalign{\kern 2pt}
&&\hskip -1cm 
A_y (x,  y + 2\pi R ) =  U A_y (x,   y) U^\dagger   ~~.
\label{BC2} 
\eeqn
Notice the relative minus sign under $Z_2$ parity reflection.

Higgs fields and bulk fermion fields must satisfy similar relations.  
Take a bulk fermion field $\psi(x,y)$.  Gauge covariance of a 
covariant derivative $D_\mu \psi$ demands that
\beqn
&&\hskip -1cm 
\psi(x,-y) = \pm   T_\psi[P_0] \gamma^5 \psi(x,y) \cr
\noalign{\kern 10pt}
&&\hskip -1cm 
\psi(x, \pi R -y) = \pm  e^{i\pi\beta_\psi} T_\psi[P_1]
  \gamma^5 \psi(x,\pi R + y)  \cr
\noalign{\kern 10pt}
&&\hskip -1cm 
\psi (x, y + 2\pi R)  = e^{i\pi \beta_\psi} T_\psi[U] \psi (x,y) 
\label{BC3}
\eeqn
where $T[P]$ represents an appropriate representation matrix.  One
immediate consequence is that a mass term $\psibar \psi$ is not
allowed on ${\mathcal M}$.  

If $P_0$ or $P_1$ is not proportional to the identity, the original
gauge symmetry is partially broken.  It gives a genuine device
to achieve gauge symmetry breaking without ``Higgs
scalar fields''.  This feature has been successfully utilized to
explain the triplet-doublet splitting problem in the $SU(5)$
model by Kawamura.\cite{Kawamura}  However, 
 at the same time it brings  about arbitrariness
or indeterminacy in the symmetry breaking pattern.  This dilemma can be
resolved by two distinct mechanisms.  The first one is to ensure that
different sets of boundary conditions $(P_0, P_1)$ lead to the same
physics, thus to make the choice of $(P_0, P_1)$ irrelevant.  The second
one is to provide dynamics to select  $(P_0, P_1)$.  In the final theory
both mechanisms most likely will come into operation. 
In this article we show that the first mecanism is indeed in action.

\section{Residual gauge invariance of the boundary conditions}

It is necessary to pin down which parts of the original gauge symmetry
are left unbroken by the orbifold boundary conditions.  Under a general
gauge transformation on the covering space $\coverM$
\beqn
A_M &\to&  {A'}_M = \Omega A_M  \Omega^{\dagger}
   -  {i \over g}\Omega  \partial_M \Omega^{\dagger} ~~, 
\label{gauge1}
\eeqn
new gauge potentials satisfy, in place of (\ref{BC1}) and (\ref{BC2}), 
\beqn
&&\hskip -1cm 
\left[ \matrix{A_\mu' (x, - y) \cr  A_y' (x, - y) \cr} \right]
= P_0' \left[ \matrix{ A_\mu' (x,y) \cr - A_y' (x,y) \cr} \right]
    P_0'^\dagger
   - {i\over g} \,  P_0' \,
  \left[ \matrix{ \dd_\mu \cr - \dd_y \cr} \right] P_0'^\dagger \cr
\noalign{\kern 10pt}
&&\hskip -1cm 
\left[ \matrix{A_\mu' (x,\pi R- y) \cr A_y' (x,\pi R- y) \cr} \right]
= P_1' \left[ \matrix{A_\mu' (x,\pi R + y) \cr -A_y' (x,\pi R + y) \cr}
   \right]  P_1'^\dagger
   - {i\over g} \,  P_1' \,
\left[ \matrix{ \dd_\mu \cr -\dd_y\cr} \right] P_1'^\dagger   \cr
\noalign{\kern 5pt}
&&\hskip -.8cm 
A_M' (x, y + 2\pi R) = U' A_M'(x,y) U'^\dagger
   - {i\over g} U' \dd_M U'^\dagger 
\label{newBC1}
\eeqn
where
\beqn
&&\hskip -1cm
P_0' = \Omega(x,-y) \, P_0 \, \Omega^\dagger (x,y) \cr
\noalign{\kern 5pt}
&&\hskip -1cm
P_1' = \Omega(x,\pi R -y) \, P_1 \, \Omega^\dagger (x,\pi R + y)  \cr
\noalign{\kern 5pt}
&&\hskip -1cm
U' = \Omega(x,y+2\pi R) \,  U \, \Omega^\dagger (x,y)  ~~.
\label{newBC2}
\eeqn
The theory, or more precisely speaking, the Hilbert space of the theory,
is defined with the orbifold boundary condtions specified.  {\bf The 
residual gauge symmetry} in the theory consists of gauge transformations
which preserve the boundary conditions so that 
\beqn
&&\hskip -1cm
\Omega(x,-y) \, P_0  = P_0 \, \Omega (x,y) \cr
\noalign{\kern 5pt}
&&\hskip -1cm
\Omega(x,\pi R -y) \, P_1  = P_1 \, \Omega (x,\pi R + y) \cr
\noalign{\kern 5pt}
&&\hskip -1cm
\Omega(x,y+2\pi R) \,  U  = U \, \Omega (x,y) ~~.
\label{residual1}
\eeqn

The residual gauge symmetry is large.  Although $(P_0, P_1, U)$
may not be invariant under global transformations of the gauge group,
$\Omega(x,y)$'s satisfying $(P_0', P_1', U') = (P_0, P_1, U)$ extend
over the whole group.

One example is in order.  Take a $SU(2)$ gauge theory with  boundary
conditions $(P_0, P_1) = (\tau_3, 
\tau_3 \cos 2\pi\alpha + \tau_1 \sin 2\pi \alpha)$.  The residual global
symmetry is $U(1)$ for $\alpha = 0, \pm \onehalf, \pm 1, \cdots$, and
none left otherwise.   The residual gauge symmetry is given by
\beqn
&&\hskip -1cm
\Omega(x,y) = \exp \bigg\{ i \sum_{a=1}^3 \omega_a(x,y) \tau_a \bigg\} \cr
\noalign{\kern 10pt}
&&\hskip -1cm
\omega_2(x,y) = \sqrt{{2\over \pi R}} \sum_{n=1}^\infty
   \omega_{2, n}(x) \sin {ny\over R} \cr
\noalign{\kern 10pt}
&&\hskip -1cm
\pmatrix{\omega_1(x,y) \cr \omega_3(x,y) \cr} =
{1\over \sqrt{\pi R}} \sum_{n=-\infty}^\infty
v_n(x) \pmatrix{ \sin \mybig{(n+2\alpha) y/ R} \cr
                  \cos \mybig {(n+2\alpha) y/ R} \cr}  ~~.
\label{gauge2}
\eeqn
These gauge transformations mix all Kaluza-Klein modes.

In many situations we are interested in physics at low energies,
or symmetry seen at an energy scale much lower than $1/R$, for which
only $y$-independent gauge transformations are recognized.  In the 
$SU(2)$ example presented above such symmetry survives for an integral 
$2\alpha$, i.e.\ the $U(1)$ gauge symmetry with $\omega_3 \sim 
v_{-2\alpha}(x)$ remains unbroken.  In general cases such low energy
gauge symmetry is given by $\Omega(x)$'s satisfying
\beqn
&&\hskip -.5cm
\Omega(x) \, P_0  = P_0 \,  \Omega (x) ~, \cr
\noalign{\kern 5pt}
&&\hskip -.5cm
\Omega(x) \, P_1 = P_1  \Omega (x) ~, \cr
\noalign{\kern 5pt}
&&\hskip -.5cm
\Omega(x) \,  U \,  = \,  U \, \Omega (x) ~,
\label{residual2}
\eeqn
that is, the symmetry is generated by generators which commute with
$P_0$, $P_1$ and $U$.  This symmetry is called  
{\bf the low energy symmetry of the boundary conditions}.

\section{Wilson line phases}

Given the orbifold boundary conditions $(P_0, P_1, U)$ there appear new
physical degrees of freedom which are absent in the Minkowski spacetime. 
Consider a path-ordered integral along a non-contractible loop on $S^1$
;
\beeq
W(x,y) =P \exp \bigg\{ ig  \int_y^{y+2\pi R} dy' \, A_y(x,y') \bigg\} 
    ~~. 
\label{wilson1}
\eneq
Under a gauge transformation $\Omega(x,y)$
\beqn
W(x,y) \, U
&\go&  \Omega(x,y) W(x,y) \Omega(x, y+2\pi R)^\dagger \, U \cr
\noalign{\kern 5pt}
&=& \Omega(x,y) W(x,y)  \, U \Omega(x, y)^\dagger
\label{wilson2}
\eeqn
where the last relation in (\ref{residual1}) has been made use of.
In other words the eigenvalues of $W(x,y)  U$ are invariant under residual 
gauge transformations. Nontrivial $(x,y)$ dependence results when 
field strengths $F_{MN} \not= 0$.  $F_{MN}=0$ does not
necessarily imply trivial $WU$, however.  

Consider a configuration with constant $A_y$ and vanishing $A_\mu$,
which certainly has $F_{MN}=0$. To satisfy the orbifold boundary conditions
(\ref{BC2}), $A_y$ must anticommutes with $P_0$ and $P_1$.  This
configuration yields $WU = \exp \{ 2\pi i g R A_y \} \cdot U$
which in general is gauge-inequivalent to $WU=1$.  Nontrivial phases
are called {\bf Wilson line phases} which are promoted to  physical 
degrees of freedom.  Let us write 
$A_M = \sum_a \onehalf A^a_M \lambda^a$ where 
$\Tr \lambda^a \lambda^b = 2 \delta^{ab}$.  Wilson line phases  on 
$M^4 \times (S^1/Z_2)$ are
$\{ \theta_a = g \pi R A_y^a ~,~ a \in \H_W \}$ where 
\beeq
\H_W = \bigg\{ ~{\lambda^a\over 2} ~;~
\{ \lambda^a, P_0 \} = \{ \lambda^a, P_1 \} =0 ~
\bigg\} ~~.
\label{wilson3}
\eneq

The presence of Wilson line phases as physical degrees of freedom reflects
the  degeneracy in  classical vacua.  The degenerate vacua are connected
by Wilson line phases.   The degeneracy is lifted by quantum
effects.  It is at this place where dynamics of Wilson line phases 
induces rearrangement of gauge symmetry.

\section{Equivalent classes of the orbifold boundary conditions}

To further motivate investigating dynamics of Wilson lines we take a
closer look at interrelations among different sets of boundary conditions. 
Recall that gauge-transformed potentials satisfy new boundary 
conditions given in Eqs.\ (\ref{newBC1}) and (\ref{newBC2}).  If 
$(P_0', P_1', U')$ turns out constant in spacetime, i.e.\
$\dd_M P_0' = \dd_M P_1' = \dd_M U' = 0$, then the new set of boundary
conditions $(P_0', P_1', U')$ is of the allowed type.  In general,
$(P_0', P_1', U')$ is distinct from $(P_0, P_1, U)$.
The low energy  symmetry of the boundary conditions are different.

When two sets of boundary conditions are related by a
boundary-condition-changing gauge transformation, the two 
sets are said to be in the same equivalent class;
\beeq
(U', P_0', P_1') \sim (U, P_0, P_1) ~~.
\label{equiv1}
\eneq
The  relation is transitive.  This defines {\bf equivalence
classes of the boundary conditions}.

It is easy to find nontrivial examples.  Take
\beeq
\Omega(x,y) = e^{ i (y+\alpha) \Lambda}
\quad \hbox{where} \quad
\{ \Lambda, P_0 \} = \{ \Lambda, P_1 \} = 0 ~,
\label{equiv2}
\eneq
which leads to
\beeq
P_0' = e^{2i\alpha \Lambda} P_0 ~~,~~
P_1' = e^{2i(\alpha + \pi R) \Lambda} P_1 ~~,~~
U' = e^{2i \pi R \Lambda} U ~~. 
\label{equiv3}
\eneq
As the reader might recognize, a boundary-condition-changing gauge
transformation has the correspondence to a Wilson line phase.

 A boundary-condition-changing gauge transformation relates
two different theories.  There is one-to-one correspondence between 
these two theories.  As they are related by a gauge transformation,
physics of the two theories must be the same. Nevertheless, the two
sets of the boundary conditions have different symmetry.  How is it 
possible for such two theories to be equivalent?  The equivalence of the
two theories is guaranteed by the  Hosotani mechanism.

\section{The Hosotani mechanism and physical symmetry}

Quantum dynamics of Wilson line phases controle the physical symmetry 
of the theory.  The mechanism is called the Hosotani mechanism which
has originally been established in gauge theories on multiply-connected
manifolds.\mycite{Hosotani1}\cite{Hosotani2}  It applies to gauge theories
on orbifolds as
well.\mycite{Quiros1}\mycite{HHHK}\mycite{Hebecker}\cite{Lim1} The only
change is that  the degrees of freedom of Wilson line phases are
restricted on orbifolds as explained in section 4.  The mechanism can be
applied to supersymmetric theories.\cite{Takenaga}  It can induce
spontaneous SUSY breaking in the gauged supergravity model.\cite{Quiros2}

{\bf The Hosotani mechanism} consists of six parts.

\noindent  (i)  Wilson line phases along non-contractible loops become
physical degrees of freedom which  cannot be gauged away.  They
parametrize  degenerate classical vacua.

\noindent (ii) The degeneracy is lifted by quantum effects, 
unless it is strictly forbidden by  supersymmetry.  
The physical vacuum is given by
the configuration of the Wilson line phases  which minimizes the
effective potential $V_\eff$. (In two or three dimensions significant
quantum fluctuations appear around the minimum of 
$V_\eff$.\mycite{Hosotani3}\cite{Hosotani4})

\noindent (iii) If the effective potential $V_\eff$ is minimized at a
nontrivial configuration of Wilson line phases, then the gauge symmetry
is spontaneously enhanced or broken  by radiative corrections.   This
part of the mechanism is sometimes called the Wilson line symmetry
breaking in the literature.   
Nonvanishing expectation values of the Wilson line phases
give masses to those gauge fields in lower dimensions whose gauge
symmetry is broken.  Some of matter fields also acquire masses.

\noindent (iv)  All zero-modes of extra-dimensional components of gauge
fields in  the broken sector of gauge group, which may exist at the
classical level,  become massive by quantum effects.

\noindent (v) The physical symmetry of the theory is determined
by the combination of the boundary conditions and the expectation values
of the Wilson line phases.  Theories in the same equivalent class of the
boundary conditions have the same physical symmetry and physics content.

\noindent (vi) The physical symmetry of the theory is mostly dictated by
the  matter content of the theory.  It does not depend on the values of
various coupling constants in the theory.

Part (v) of the mechanism is of the biggest relevance in our discussions. 
It tells us that  the physics is independent of the orbifold boundary
conditions so long as they belong to the same equivalent class of
the boundary conditions.

The physical symmetry of the theory, $H_{\rm sym}$, is determined as
follows.  Suppose that with the boundary conditions $(P_0, P_1, U)$ 
the effective potential is minimized at $\la A_y \ra$ such that
$W = \exp ( ig2\pi R \la A_y \ra ) \not= 1$.   One needs to know the 
symmetry around $\la A_y \ra$.  Perform a boundary-condition-changing
gauge trasformation $\Omega(y) = \exp\{ ig(y+\beta) \la A_y \ra \}$,
which brings $\la A_y \ra$ to $\la A_y' \ra =0$.  At the same time the
orbifold boundary conditions change to
\beeq
(P_0^\sym, P_1^\sym, U^\sym) 
= (e^{2ig\beta \langle A_y \rangle} P_0, 
  e^{2ig(\beta +\pi R)\langle A_y \rangle} P_1, W U) 
\label{sym1}
\eneq
where we have made use of 
$\{ \la A_y \ra , P_0 \} = \{ \la A_y \ra , P_1 \} =0$.
The physical symmetry is the symmetry of 
$(P_0^\sym, P_1^\sym, U^\sym)$ as the expectation values of
$A_y'$ vanish.  In particular, {\bf the physical symmetry at low energies}
is spanned by the generators in
\beeq
\H^\sym =  \bigg\{ ~{\lambda^a\over 2} ~;~
[ \lambda^a, P_0^\sym ] = [ \lambda^a, P_1^\sym ] =0 ~ \bigg\} ~~.
\label{sym2}
\eneq
The symmetry $H^\sym$ generated by ${\mathcal H}^\sym$ does not depend on
the parameter $\beta$.

\section{Effective potential}

To find $\la A_y \ra$ it is necessary to evaluate the effective potential
for Wilson line phases.  The effective potential is most elegantly
evaluated  in the background field gauge.\cite{Hosotani2}  The effective
potential for a configuration $A_M^{0}$ is found by  writing 
$A_M = A_M^{0} + A_M^{q}$, taking 
$F[A] = D_M(A^{0})  A^{qM} = \dd_M A^{qM} + ig [A^{0}, A^{qM}] =0$
as a gauge fixing condition, and integrating over the quantum part 
$A_M^{q}$.

The effective potential in the background field gauge provides a natural
link among theories with different sets of orbifold boundary conditions.
Suppose that a gauge transformation (\ref{gauge1})  satisfies the relation
\beeq
 \dd^M (\dd_M \Omega^{\dagger} \Omega) 
  + i g[A^{0 M},  \dd_M  \Omega \Omega^{\dagger}] = 0  ~~.
\label{Veff1}
\eneq
Then it is shown that 
\beeq
V_\eff [A^{0}; P_0, P_1, U]
= V_\eff [ A'^{0} ; P_0', P_1', U'] 
\label{Veff2}
\eneq
where $(P_0', P_1', U')$ is given by (\ref{newBC2}).  As observed in
section 5, a Wilson line phase and a boundary-condition-changing gauge
transformation  have correspondence between them. 
For such  $A^{(0)}$ and $\Omega$ the relation (\ref{Veff1}) is satisfied.
The property (\ref{Veff2}) in turn implies that the minimum of the 
effective potential corresponds to the same symmetry as that of
$(P_0^\sym, P_1^\sym, U^\sym)$ in the previous section.   This establishes
the part (v) of the Hosotani mechanism.  We shall see it in more detail 
in the $SU(5)$ models in sections 8 and 9.

The one-loop effective potential  is 
given, in ${\mathcal M}$,  by
\beeq
V_\eff [A^0] = \sum \mp \,
   {i \over 2} \, \Tr \ln  D_{M}(A^0) D^{M}  (A^0) 
\label{Veff3}
\eneq
where the sum extends over all degrees of freedom of fields defined on the
bulk ${\mathcal M}$. The sign is negative (positive) for bosons (ghosts and
fermions). $D_M (A^0)$ denotes an appropriate covariant derivative with a
background field $A_M^0$.   $V_\eff$ depends on $A_M^0$ and the boundary
conditions $(P_0, P_1, U)$.

We are interested in the $A^0$-dependent part of $V_\eff$.  For a given
$A^0$ and $(P_0, P_1, U)$, one can always take a basis of fields such 
that ``$\Tr \ln$'' in (\ref{Veff3}) decomposes into singlets and doublets
of fields, among which only doublet fields yield $A^0$-dependence.
This seems to result from the nature of the $Z_2$-orbifolding.

A fundamental $Z_2$-doublet $\phi^t = (\phi_1,\phi_2)$ satisfies the
orbifold boundary conditions of the form
\beqn
&&\hskip -1cm
\phi(x,-y) = \tau_3 \phi(x,y) ~~, \cr
\noalign{\kern 5pt}
&&\hskip -1cm
\phi(x,y+2\pi R) = e^{-2\pi i\alpha \tau_2} \, \phi(x,y) ~,
\label{rep1}
\eeqn
being expanded as
\beeq
\left[ \matrix{\phi_1(x,y) \cr \phi_2(x,y) \cr} \right] =
{1\over \sqrt{\pi R}} \sum_{n=-\infty}^\infty
\phi_n(x) \left[ \matrix{ \cos \mybig{(n+\alpha) y / R} \cr
              \sin \mybig {(n+\alpha) y / R} \cr} \right] ~~.
\label{expansion1}
\eneq
The coupling of $\phi$ to Wilson line phases  is cast in the form
\beeq
\onehalf \big| D_y(A_y) \phi \big|^2 =
 \onehalf \Big(\dd_y \phi_1 - {\gamma\over R} \phi_2 \Big)^2
+ \onehalf \Big(\dd_y \phi_2 + {\gamma\over R} \phi_1 \Big)^2
\label{pair1}
\eneq
where $\gamma$ is a linear combination of Wilson line phases.
Insertion of (\ref{expansion1}) into (\ref{pair1}) yields
\beqn
&&\hskip -1cm
\int_0^{\pi R} dy \, \onehalf \big| D_y(A_y) \phi \big|^2
=  \onehalf \sum_{n=-\infty}^\infty 
{(n+\alpha +\gamma)^2\over R^2} \, \phi_n(x)^2 ~~.
\label{pair2}
\eeqn
Notice that the number of degrees of freedom is halved due to the 
$Z_2$-orbifolding compared with that on $S^1$.  
Hence the contribution of a bosonic $Z_2$-doublet $\phi$ to $V_\eff$ is
\beqn
&&\hskip -1cm 
-{i \over 2} \int {d^4 p \over (2 \pi)^4}
  ~{1 \over 2 \pi R} ~ 
 \sum_{n = -\infty}^{\infty} \ln
  \bigg\{ - p^2 +  \Big( {n+\alpha +\gamma \over R} \Big)^2 \bigg\} \cr
\noalign{\kern 5pt}
&&\hskip +1.3cm 
= - {1\over 64 \pi^7 R^5} \, f_5 \big[ 2(\alpha+\gamma) \big] 
   + \hbox{constant} 
\label{Veff4}
\eeqn
where $f_D(x) = \sum_{n=1}^\infty n ^{-D} \cos (n\pi x)$.

\section{Physical symmetry in the non-supersymmetric $SU(5)$ model}

Consider the non-supersymmetric $SU(5)$ gauge theory.   We assume that
the gauge fields and $N_h$  Higgs field in {\bf 5} live in
the bulk five-dimensional spacetime ${\mathcal M}$.  Quarks and leptons are
supposed to be confined on the boundary at $y=0$. There may be additional
$N_f^5$ and $N_f^{10}$ fermion multiplets in {\bf 5} and {\bf 10}
 defined in the bulk ${\mathcal M}$.   

Let us focus on the following boundary conditions.
\beqn
&&\hskip -1cm
(\BC 0):~
P_0 =  \diag (-1,-1,-1,1,1) ~~,~~
P_1 =  \diag (1,1,1,1,1)  \cr
\noalign{\kern 3pt}
&&\hskip -1cm
(\BC 1):~
P_0 =  \diag (-1,-1,-1,1,1) ~~,~~
P_1 =  \diag (-1,-1,-1,1,1)  \cr
\noalign{\kern 3pt}
&&\hskip -1cm
(\BC 2):~
P_0 =  \diag (-1,-1,-1,1,1) ~~,~~
P_1 =  \diag (-1,1,-1,1,-1)  \cr
\noalign{\kern 3pt}
&&\hskip -1cm
(\BC 3):~
P_0 =  \diag (-1,-1,-1,1,1) ~~,~~
P_1 =  \diag (1,1,-1,-1,-1)  \cr
\noalign{\kern 3pt}
&&\hskip -1cm
(\BC 4):~
P_0 =  \diag (-1,-1,-1,1,1)  ~~, \cr
\noalign{\kern 3pt}
&&\hskip .4cm 
P_1 =  \pmatrix{ - \cos \pi p & 0& 0& -i \sin \pi p& 0\cr
                   0& - \cos \pi q & 0& 0& -i \sin \pi q\cr
                   0& 0& -1 & 0 & 0\cr
                  i \sin \pi p& 0& 0 &    \cos \pi p & 0\cr
                    0& i \sin \pi q & 0& 0& \cos \pi q\cr} .
\label{boundary1}
\eeqn
(BC1), (BC2), (BC3) are special cases of (BC4).
Their low energy symmetry of boundary conditions is
\beqn
&&\hskip -1cm
G_\BC^{(0)} = SU(3) \times SU(2) \times U(1) \cr
\noalign{\kern 3pt}
&&\hskip -1cm
G_\BC^{(1)} = SU(3) \times SU(2) \times U(1) \cr
\noalign{\kern 3pt}
&&\hskip -1cm
G_\BC^{(2)} = SU(2) \times U(1) \times U(1) \times U(1) \cr
\noalign{\kern 3pt}
&&\hskip -1cm
G_\BC^{(3)} = SU(2) \times SU(2) \times U(1) \times U(1) \cr
\noalign{\kern 3pt}
&&\hskip -1cm
G_\BC^{(4)} = U(1) \times U(1) \times U(1) ~~.
\label{boundary2}
\eeqn

The boundary conditions (BC0) and (BC1), at a first look,  seem natural to
incorporate the standard model symmetry at low energies and to 
provide a solution to the triplet-doublet splitting problem.
Indeed, (BC0) is the orbifold conditiion originally adopted by
Kawamura.\cite{Kawamura}

One might ask why one should take (BC0) or (BC1)?  Why can't we adopt
(BC2), (BC3), or even (BC4)?  We shall demonstrate that, if (BC1) is a
legitimate choice for the boundary conditions, then (BC2), (BC3), and 
(BC4) are as well.  It does not matter which one to
choose, as all of them lead to the same physics by the Hosotani mechanism.

First we note that the equivalence class of boundary conditions to which
(BC0) belongs consists of only one element, namely (BC0) itself.
There is no Wilson line phase in the theory with (BC0), as $P_1$ is 
the identity.

(BC1), (BC2), (BC3) and (BC4) belong to the same equivalence class
of boundary conditions.  (BC1) and (BC4) are related to each other by
a boundary-condition-changing gauge transformation
\beeq
\Omega (y)  = \exp \big\{ - i(y /2 R) T_{p,q} \big\} ~~,~~
T_{p,q} = \pmatrix{ 0 & 0 & 0 & p & 0\cr
                0 & 0 & 0 & 0 & q\cr
                0 & 0 & 0 & 0 & 0\cr
                p & 0 & 0 & 0 & 0\cr
                0 & q & 0 & 0 & 0\cr}  ~~.
\label{equiv4}
\eneq
Hence all of (BC1) to (BC4) should have the same physics, which is
confirmed by explicit computations of the effective potential for Wilson 
line phases.

In the theory with (BC1), Wilson line phases are given by
\beeq
2g\pi R A_y = \pi 
      \pmatrix{ 0 & 0 & 0 & c_1 & c_4\cr
                0 & 0 & 0 & c_2 & c_5\cr
                0 & 0 & 0 & c_3 & c_6\cr
                c_1^* & c_2^* & c_3^* & 0 & 0\cr
                c_4^* & c_5^* & c_6^* & 0 & 0\cr} ~~.
\label{wilson4}
\eneq
There are twelve Wilson line phases.  In the theory with (BC4), however,  
there survive only two phases;
\beeq
2g\pi R A_y = \pi  \, T_{a,b}
\label{wilson5}
\eneq
where $(a,b)$ are real.  In evaluating the effective potential for 
(\ref{wilson4}) in the theory (BC1), one can utilize the residual 
$SU(3) \times SU(2) \times U(1)$ invariance to reduce (\ref{wilson4})
to (\ref{wilson5}).  Hence it is sufficient to evaluate the effective
potential $V_\eff^{(p,q)} (a, b)$ 
for the configuration (\ref{wilson5}) in the theory (BC4)
which includes (BC1) as a special case $(p,q) = (0,0)$.

The evaluation is straightforward.  It is reduced to
identifying  all $Z_2$ singlets and doublets as described in section 7. 
The result is  
\beqn
&&\hskip -1cm
V_\eff^{(p,q)} (a,b) = - {3\over 64\pi^7 R^5} 
\Big\{ N_A \big[ f_5(a-p)+ f_5(b-q) \big]  \cr
\noalign{\kern 3pt}
&&\hskip 2.8cm
+N_B  \big[ f_5(a+b-p-q) + f_5(a-b-p+q) \big] \cr
\noalign{\kern 3pt}
&&\hskip 2.8cm
+{3 \over 2} \big[ f_5(2a-2p) + f_5(2b-2b) \big] ~ \Big\}
\label{Veff5}
\eeqn
where
$N_A \equiv 3+N_h-2N_f^5-2N_f^{10}$ and  $N_B \equiv 3-2N_f^{10}$.
There are a few features to be noted;   
(i) $V_\eff^{(p,q)} (a,b) = V_\eff^{(q,p)} (b,a)$.  
(ii) $V_\eff$ is  periodic in $(a,b)$ with a period 2.  
(iii) $V_\eff^{(p,q)} (a,b) = V_\eff^{(0,0)} (a-p,b-q)$.
(iv) The form of $V_\eff$ and the location
of its minimum are determined by $N_h$, $N_f^5$, and $N_f^{10}$, namely by
the matter content.  

The fact that $V_\eff$ is a function of $a-p$ and $b-q$ is of critical
importance.  It manifests the relation (\ref{Veff2}), implying that 
the physical symmetry determined by the minimum of $V_\eff$ is 
independent of $(p,q)$.  The minimum is located at $(a-p,b-q)=(0,0)$,
$(1,1)$, or $(0,1)\sim (1,0)$, depending on the matter content.

Some examples are in order.  In figure 1 $V_\eff^{(0,0)} (a,b)$ is plotted
for various  $(N_h, N_f^5, N_f^{10})$; (a) (1,0,0)$, 
(b) (1,3,3)$, (c) (1,1,1), and (d) (1,0,2).  The minimum is located at
(a) $(a,b)=(0,0)$, (b) $(1,1)$, (c) (0,0), and (d) (0,1).  In the case (c)
(0,0) and (1,1) are almost degenerate. 

\begin{figure}[tbh]
\leavevmode
\mbox{
\epsfysize=4.2cm \epsfbox{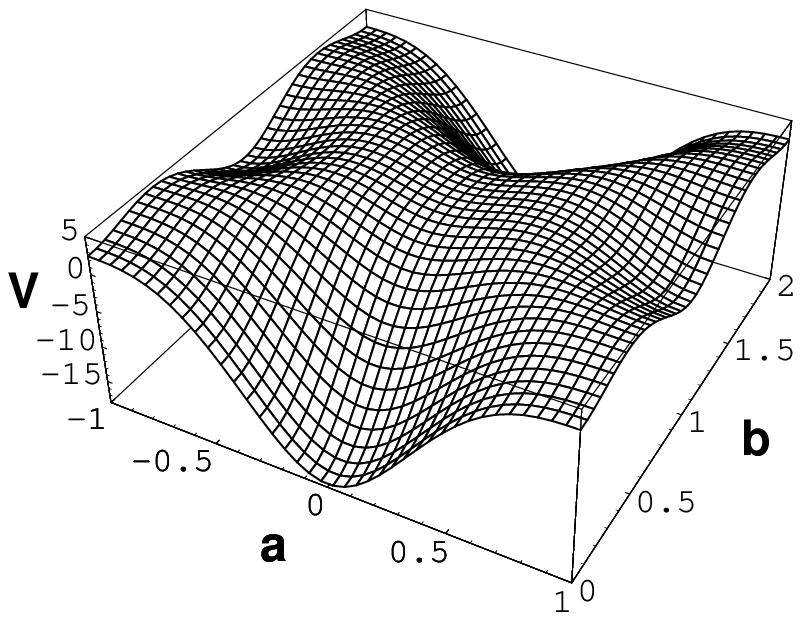}}
\hskip -.2cm
\mbox{
\epsfysize=4.2cm \epsfbox{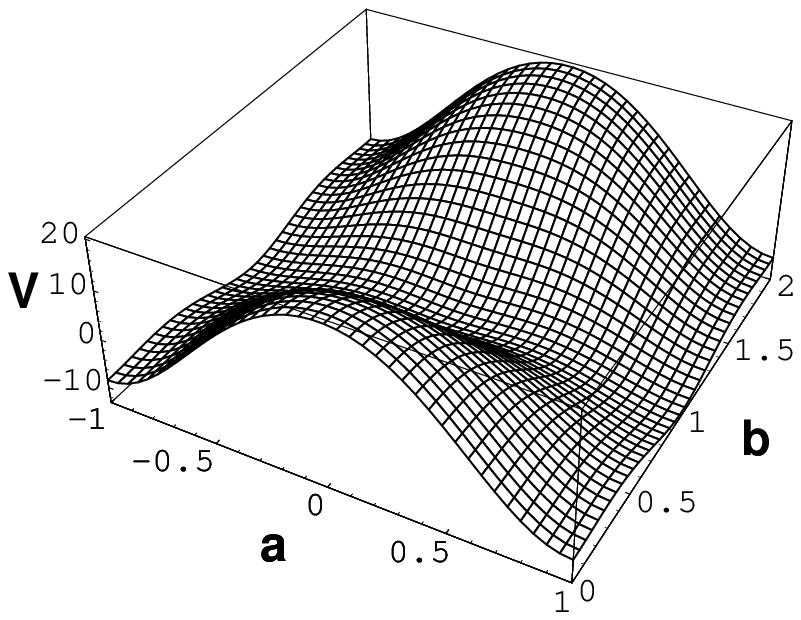}}

\vskip -.5cm
\hskip .5cm (a) ~ (1,0,0) \hskip 3.5cm (b) ~ (1,3,3)

\mbox{
\epsfysize=4.2cm \epsfbox{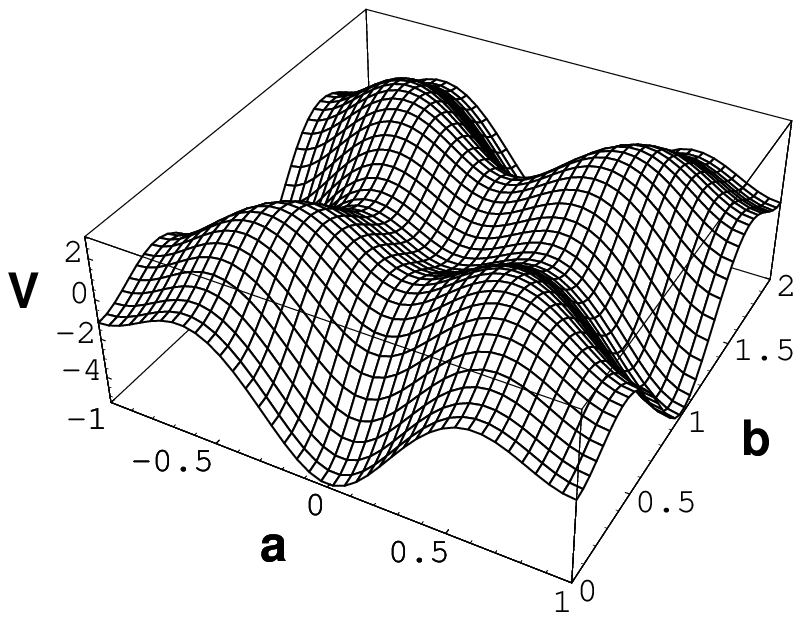}}
\hskip -.2cm
\mbox{
\epsfysize=4.2cm \epsfbox{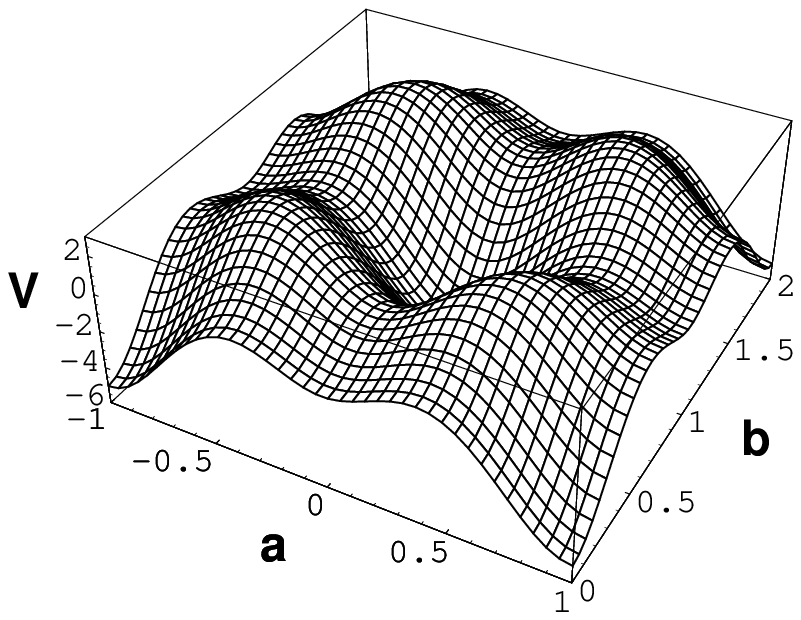}}

\vskip -.5cm
\hskip .5cm (c) ~ (1,1,1) \hskip 3.5cm (d)  ~ (1,0,2)

\caption{The effective potential for various $(N_h,  N_f^5, N_f^{10})$
in the non-supersymmetric models.  $V_\eff(a,b)/C$ ($C=3/64 \pi^7 R^5$) is
plotted for $(p,q)=(0,0)$ in (\ref{Veff5}).}
\label{fig-Veff1}
\end{figure}

The physical symmetry at low
energies in the case (a)  is 
$H^\sym = G_\BC^{(1)} = SU(3) \times SU(2) \times U(1)$.  
In the case (b) we recall that $(a,b; p,q)=(1,1;0,0)$ 
is equivalent to $(a,b; p,q)=(0,0;1,1)$.  Hence the physical symmetry is
$H^\sym = G_\BC^{(3)} = [SU(2)]^2  \times [U(1)]^2$.  In the case (d)
$H^\sym = G_\BC^{(2)} = SU(2)  \times [U(1)]^3$.

The resultant physical symmetry is independent of the values of 
$(p,q)$ in the boundary conditions.  It is determined solely by
the matter content in the bulk ${\mathcal M}$.  Dynamical rearrangement of 
gauge symmetry has taken place as a result of quantum dynamics of 
the Wilson line phases.   Symmetry can be spontaneously
enhanced or broken, depending on the boundary conditions.

\section{Physical symmetry in the supersymmetric $SU(5)$ model}

If the theory has supersymmetry which remains unbroken, then the 
effective potential for Wilson lines stays flat due to the 
cancellation among contributions from bosons and fermions.
Nontrivial dependence in $V_\eff$ appears if the supersymmetry is
softly broken as the nature demands.

There is a natural way to introduce soft SUSY breaking on 
multiply connected manifolds and orbifolds. First note that 
$N=1$ SUSY in five dimensions induces $N=2$ SUSY in four dimensions.
A five-dimensional (5-D) gauge multiplet 
${\mathcal{V}}=(A_M, \lambda, \lambda', \sigma)$ is decomposed to a vector
superfield $V=(A_\mu, \lambda)$ and a chiral superfield
$\Sigma = (\sigma + iA_y, \lambda')$ in four dimensions.
Similarly, 5-D fundamental Higgs hypermultiplets 
${\mathcal H}=(h, h^c{}^\dagger, \tilde{h},   \tilde{h}^c{}^\dagger)$ and 
$\overline{{\mathcal H}}=(\overline{h}, \overline{h}^c{}^\dagger,
   \tilde{\overline{h}}, \tilde{\overline{h}}^c{}^\dagger)$ are
decomposed into 4-D chiral superfields
$H=(h, \tilde{h})$, $\overline{H}=(\overline{h}, \tilde{\overline{h}})$,
$H^c=(h^c, \tilde{h}^c)$, and 
$\overline{H^c}=(\overline{h}^c, \tilde{\overline{h}^c})$.  
After a  translation along a noncontractible loop, fields may have
different twist, depending on their $SU(2)_R$ charges.
This is called the Scherk-Schwarz SUSY breaking.\cite{SS}

On the orbifold ${\mathcal M}$ this twisting is implemented for
$SU(2)_R$ doublets by generalizing
(\ref{rep1}).\mycite{Quiros3}\cite{Barbieri1}  It reads, for the gauge
multiplet ${\mathcal{V}}$, that
\beqn
&&\hskip -1cm
\pmatrix{ V \cr \Sigma\cr} (x, -y)
= P_0 ~\pmatrix{ V \cr -\Sigma\cr}  (x, y) ~ P_0^\dagger \cr
\noalign{\kern 5pt}
&&\hskip -1cm
\pmatrix{A_M \cr \sigma \cr}  (x^\mu, y+2 \pi R)
  = U ~ \pmatrix{A^M \cr \sigma \cr}  (x^\mu, y) ~ U^\dagger ~~, \cr
\noalign{\kern 5pt}
&&\hskip -1cm
\pmatrix{  \lambda \cr \lambda' \cr} (x, y+2\pi R)
= e^{-2\pi i \beta \sigma_2} ~
U ~\pmatrix{  \lambda \cr \lambda' \cr}  ~U^\dagger ~~.
\label{SS1}
\end{eqnarray}
Similarly nontrivial twisting is imposed on $(h, h^{c \dagger})$ and 
$({\overline h}, {\overline h}^{c \dagger})$ doublets.  The Scherk-Schwarz
parameter $\beta$ changes the spectrum, giving rise to the SUSY breaking
scale
$M_{\rm SUSY} \sim \beta/\pi R$.  

The effective potential for Wilson line phases  for the theory with $N_h$
sets of Higgs hypermultiplets ${\mathcal H}+\overline{\mathcal H}$ is
\beqn
&&\hskip -.7cm 
V_\eff^{(0,0)} (a,b)
= - {3\over 32 \pi^7 R^5} \sum_{n = 1}^{\infty}
{1-\cos 2\pi n\beta \over n^5} 
~ \Big\{ 2(1-2N_h)(\cos \pi na +\cos \pi nb) \cr
\noalign{\kern 5pt}
&&\hskip 2.4cm
 +4\cos \pi na  \cos \pi nb   +\cos 2\pi na +\cos 2\pi nb \Big\} ~.
\label{SUSYeffpot}
\eeqn
It vanishes at $\beta=0$.  The Higgs multiplets significantly affect the
shape of the effective potential, and consequently the physical symmetry
of the theory. 

\begin{figure}[tbh]
\leavevmode
\mbox{
\epsfysize=4.2cm \epsfbox{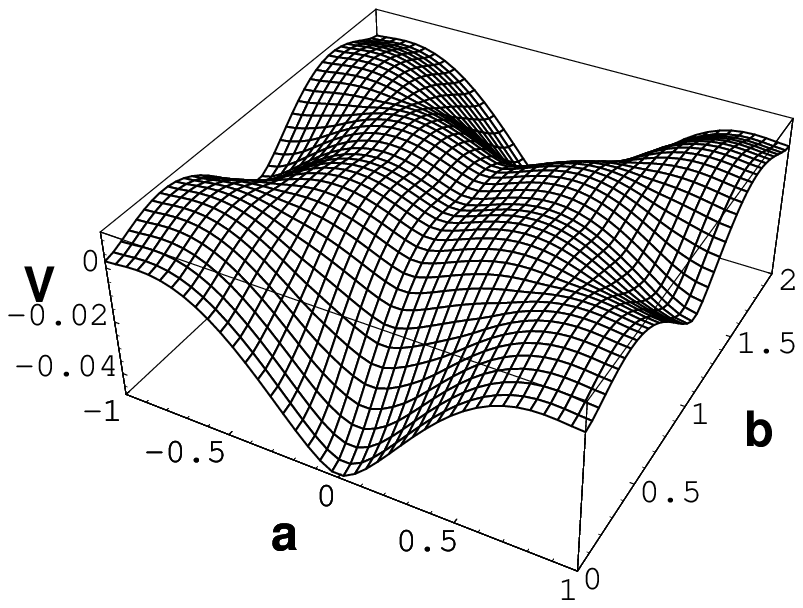}}
\hskip -.2cm
\mbox{
\epsfysize=4.2cm \epsfbox{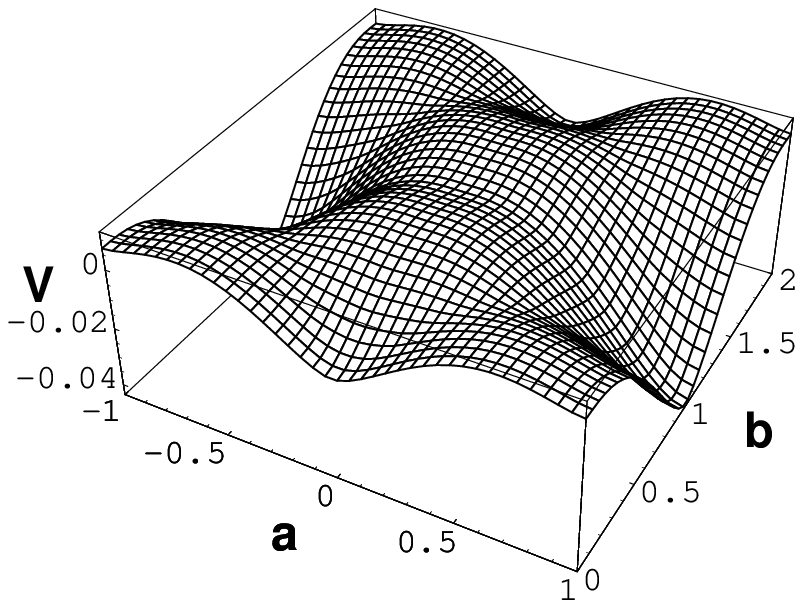}}

\vskip -.5cm
\hskip .8cm (a) ~ $N_h=0$  \hskip 3.5cm (b) ~ $N_h = 1$

\caption{The effective potential in the supersymmetric models.
$V_\eff(a,b)/C$ is plotted.  
The figures are given for the Scherk-Schwarz parameter $\beta=0.01$.}
\label{fig-Veff2}
\end{figure}

$V_\eff$ is plotted in figure 2 with the boundary conditions (BC1) for (a)
$N_h=0$ and (b) $N_h = 1$. 
If there is no Higgs multiplet, then $V_\eff$ is minimized at $(a,b)=(0,0)$
so that the physical symmetry is $H^\sym = G_\BC^{(1)}$.  For $N_h \ge 1$, 
 $V_\eff$ is minimized at $(a,b)=(1,1)$
so that the physical symmetry is $H^\sym = G_\BC^{(3)}$.
The presence of Higgs multiplets induces the breaking of color $SU(3)$
down to $SU(2) \times U(1)$.  

As stated in part (iv) of the Hosotani mechanism in section 6, all
extra-dimensional components of gauge fields in  the broken sector of
gauge group become massive by quantum effects.  The magnitude of
their masses is $g_4/R \sim g_4 M_{\rm GUT}$ in the non-supersymetric
models, while $g_4 \beta/R \sim g_4 M_{\rm SUSY}$ in the supersymmetric
models.
 
\section{Summary}

In gauge theory on orbifolds boundary conditions have to be specified.
The arbitrariness problem in the choice of the orbifold boundary
conditions is partially solved by the Hosotani mechanism.  Various sets
of boundary conditions are related by boundary-condition-changing
gauge transformations, thus falling in one equivalent class of the boundary
conditions.  Theories in the same equivalent class, though they in general
have different symmetry of boundary conditions at the classical level, 
have the same physics thanks to the dynamics of Wilson line phases.
The physical symmetry is determined by the matter content of the theory

\begin{figure}[tbh]
\centering
\leavevmode
\mbox{
\epsfysize=5.0cm \epsfbox{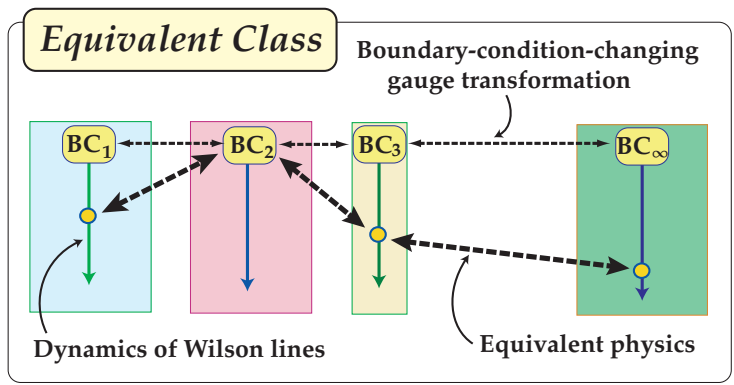}}
\caption{The concept of the Hosotani mechanism on orbifolds.  
Physics is the same in all theories in one equivalent
class of boundary conditions.  In this example the physical symmetry is
the symmetry of the boundary condition BC$_2$,  irrespective of the 
boundary condition chosen.  Dynamics of Wilson line phases brings the
true vacuum to the configuration depicted as a small circle in each
theory.}
\label{fig-concept}
\end{figure}
 
The concept is schematically depicted in figure 3.  In each theory with
given boundary condition BC$_a$ there appear degrees of freedom of  Wilson
line phases.  Their dynamics selects a particular configuration of the
Wilson line phases which minimizes the effective potential and defines
the physical vacuum.  The selected configuration always has the same
physics content, independent of the boundary condition BC$_a$.  All of
these have been confirmed in the various $SU(5)$ models.

There are several things to be investigated.

\noindent
[1] ~We need to classify all equivalent classes of boundary
condition. This poses an interesting mathematical exercise.

\noindent
[2] ~We have shown that physics is the same in each equivalent class,
but we have not so far explained which equivalent class one should start
with. It is most welcome to have a dynamical mechanism to select an
equivalent class of boundary conditions.

\noindent
[3] ~The models discussed in this paper is not entirely realistic.
For instance, we have not implemented the electroweak symmetry breaking
and quark-lepton masses.  

\noindent
[4] ~The fundamental Higgs field has not been unified with gauge fields.
Their coupling to quarks and leptons remain arbitrary.

We shall come back to these points in due course.

\section*{Acknowledgement}

This work was supported in part by  Scientific Grants
from the Ministry of Education and Science, Grant No.\ 13135215 and 
Grant No.\ 13640284.

\def\jnl#1#2#3#4{{#1}{\bf #2} (#4) #3}

\def\Zphys{{\em Z.\ Phys.} }
\def\jssc{{\em J.\ Solid State Chem.\ }}
\def\jpsJ{{\em J.\ Phys.\ Soc.\ Japan }}
\def\ptps{{\em Prog.\ Theoret.\ Phys.\ Suppl.\ }}
\def\PTP{{\em Prog.\ Theoret.\ Phys.\  }}

\def\JMP{{\em J. Math.\ Phys.} }
\def\NPB{{\em Nucl.\ Phys.} B}
\def\NP{{\em Nucl.\ Phys.} }
\def\PLB{{\em Phys.\ Lett.} B}
\def\PL{{\em Phys.\ Lett.} }
\def\PRL{\em Phys.\ Rev.\ Lett. }
\def\PRB{{\em Phys.\ Rev.} B}
\def\PRD{{\em Phys.\ Rev.} D}
\def\PRe{{\em Phys.\ Rep.} }
\def\AP{{\em Ann.\ Phys.\ (N.Y.)} }
\def\RMP{{\
em Rev.\ Mod.\ Phys.} }
\def\ZPC{{\em Z.\ Phys.} C}
\def\SCI{\em Science}
\def\CMP{\em Comm.\ Math.\ Phys. }
\def\MPLA{{\em Mod.\ Phys.\ Lett.} A}
\def\IJMPA{{\em Int.\ J.\ Mod.\ Phys.} A}
\def\IJMPB{{\em Int.\ J.\ Mod.\ Phys.} B}
\def\EPJC{{\em Eur.\ Phys.\ J.} C}
\def\PR{{\em Phys.\ Rev.} }
\def\JHEP{{\em JHEP} }
\def\cmp{{\em Com.\ Math.\ Phys.}}
\def\JPA{{\em J.\  Phys.} A}
\def\CQG{\em Class.\ Quant.\ Grav. }
\def\ATMP{{\em Adv.\ Theoret.\ Math.\ Phys.} }
\def\ibid{{\em ibid.} }

\end{document}